\documentclass[twoside]{dis04}
\def\runauthor{Joakim Nystrand}
\def\shorttitle{Deuteron and anti-deuteron production}
\begin{document}

\title{Results from PHENIX on deuteron and anti-deuteron production in Au+Au 
Collisions at RHIC}

\author{
Joakim Nystrand \\ for the PHENIX Collaboration
\footnotemark
}
\footnotetext{*See \cite{Adler:2004uy} for the full PHENIX author and institution list.}

\address{Department of Physics and Technology \\
University of Bergen, All\'{e}gaten 55, 5007 Bergen, Norway \\
E-mail: Joakim.Nystrand@ift.uib.no}

\maketitle

\abstracts{
Results from the PHENIX Collaboration on the production of 
deuterons and anti-deuterons in collisions between gold nuclei 
at a nucleon-nucleon center-of-mass energy of $\sqrt{s_{NN}} =$~200~GeV 
are presented.
}

At the Relativistic Heavy Ion Collider (RHIC) at Brookhaven National Laboratory, protons and 
gold nuclei have been brought to collide at nucleon-nucleon center-of-mass energies up to 
$\sqrt{s_{NN}} =$~200~GeV. The data this presentation is based on are from Au+Au interactions at 
the maximum RHIC energy and were accumulated during the second run at RHIC in the year 2001. 
The results on deuteron and anti-deuteron production from this run have already been submitted 
for publication\cite{Adler:2004uy}. In this talk, these results will be summarized and some of 
the experimental details not covered in the submitted publication will be discussed. 

The PHENIX experiment is a versatile system of detectors aimed at studying a variety of 
observables in heavy ion interactions. The main parts are two central tracking arms, centered 
around mid-rapidity, and two muon arms at forward/backward rapidities. Each central tracking 
arm covers $\Delta \phi$~=~90$^o$ in azimuth and $| \eta | \leq$~0.35 in pseudorapidity 
($\eta = \ln ( \tan(\Theta/2) )$, where $\Theta$ is the emission angle). Zero-degree calorimeters 
(ZDC) and Beam-beam Cherenkov counters (BBC) are used for triggering and event selection. Further 
details about PHENIX can be found in \cite{Adcox:2003zm}.

The detectors used for this analysis are, in addition to the BBCs and ZDCs, the drift- and 
pad-chambers (DC and PC) and the Time-of-Flight detector (ToF) from one of the central arms. 
Tracking and momentum reconstruction of charged particles is provided by the information 
from the DC and PC. The flight-time measurement from the ToF is combined with the starting 
time (t0) from the BBC to determine the particle velocity. Figure 1 (left) shows a 
scatter plot of $q \cdot (1-\beta)$ vs. $1/p$ for reconstructed tracks. Here, $q$ is the electric 
charge of the reconstructed particle in units of $e$, $\beta$ the velocity in units of $c$, and 
$p$ the reconstructed momentum. In addition to pion, kaons and protons, bands corresponding to 
deuterons and anti-deuterons can easily be identified. From $\beta$ and $p$, the particle mass, $m$,  
can be calculated; Figure 1 (right) shows the distribution for positive and negative particles. 

The yields of deuterons and anti-deuterons are obtained by fitting the $m^2$ distribution to 
the sum of a gaussian (representing the signal) and an exponential (representing the background) 
function, as described in \cite{Adler:2004uy}. The fits were performed in 7 bins in $p_T$ in the 
range 1.1~$\leq p_T \leq$~4.3. The total number of reconstructed deuterons and anti-deuterons for 
three centrality selections is listed in Table~1. 

\begin {table} [hbt] \begin{center} 
\begin{tabular} { c | r | r | r | r } \hline 
Centrality & Number of & 
$<N_{part}>$  & \multicolumn{2}{c}{Reconstructed} \\ 
              & \multicolumn{1}{|c|}{Events} & & deuterons & anti-deuterons \\ \hline
Min. Bias  &  $21.6 \cdot 10^6$ & 109.1$\pm$4.1  &  3140 &  1510 \\
0-20 \%    &  $4.7  \cdot 10^6$ & 279.9$\pm$4.0  &  1850 &   870 \\
20-92 \%   &  $16.9 \cdot 10^6$ & 61.9$\pm$3.2   &  1330 &   650 \\ \hline
\end{tabular}
\label{Table:yields}
\caption{The number of events, the mean number of participants, and the number of reconstructed deuterons 
and anti-deuterons for three centrality selections. The number of 
reconstructed \mbox{(anti-)}deuterons was obtained from separate fits for each centrality 
selection and $p_T$ bin. 
This is the reason why the numbers for the subsamples do not add up exactly to the numbers 
for the min. bias sample. The yields are consistent if the errors from the fits are taken into account.} 
\vspace{-0.7cm}
\end{center} \end{table}

The corrections needed to convert the raw yields to normalized deuteron/anti-deuteron spectra 
are obtained by running simulated 
tracks through a GEANT 3.21 based Monte Carlo simulation program of the detector\cite{Adler:2003cb}. 
For hadrons, this simulation includes particle decay (if applicable), 
multiple Coulomb scattering, hadronic interactions, and the acceptance, efficiency and resolution of 
the detector. 

The hadronic interactions in Geant 3.21 are handled by FLUKA\cite{Fasso:2000hd}. 
Although the implementation of nucleus-nucleus interactions in FLUKA is in 
progress\cite{Fasso:2003xz}, the version used by Geant 3.21 does not include 
such processes. The correction for hadronic interactions of (anti-)deuterons 
thus has to be implemented separately. 

\begin{figure}[!b]
\begin{center}
\centerline{\epsfxsize=2.2in\epsfbox{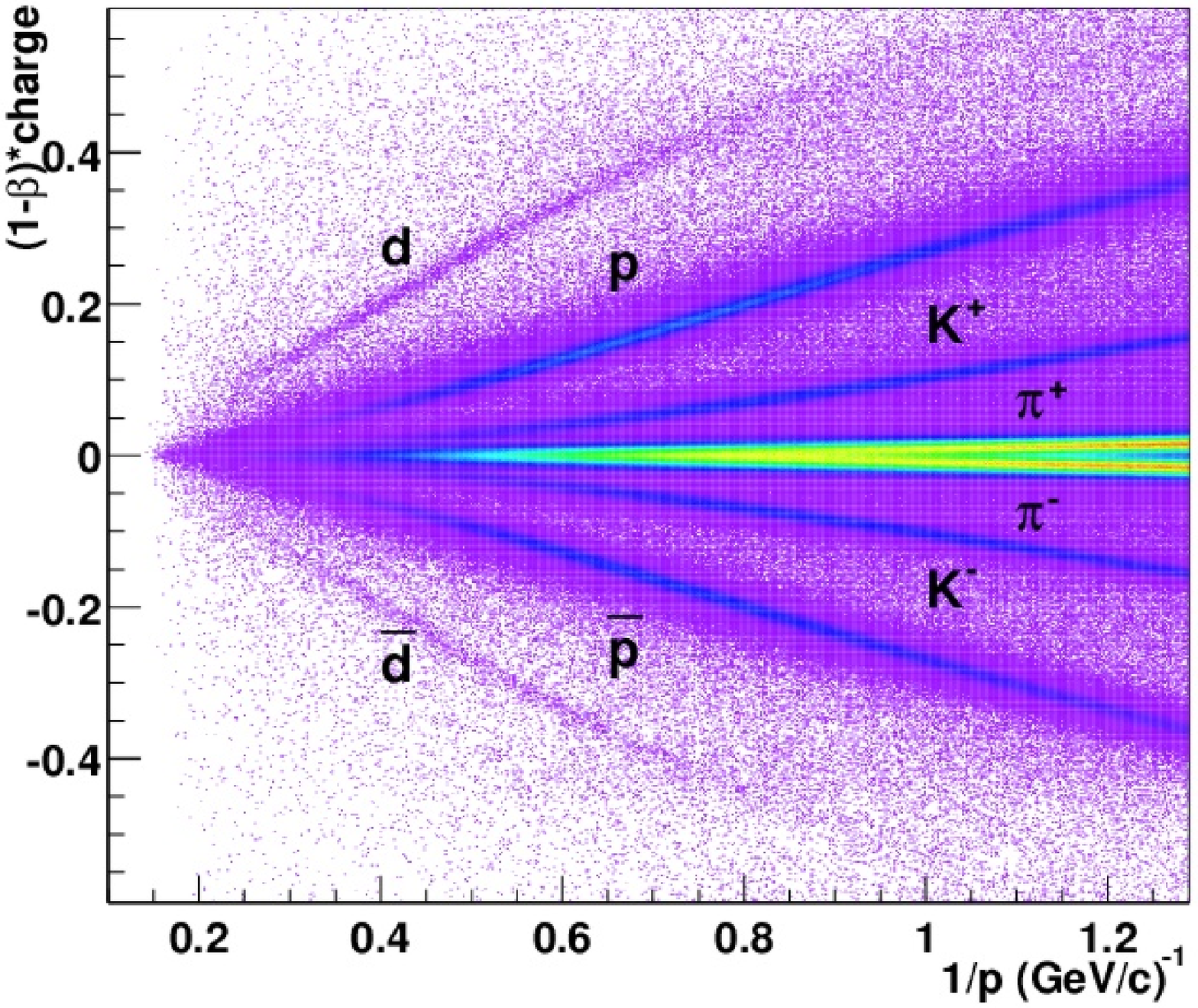} \epsfxsize=2.2in\epsfbox{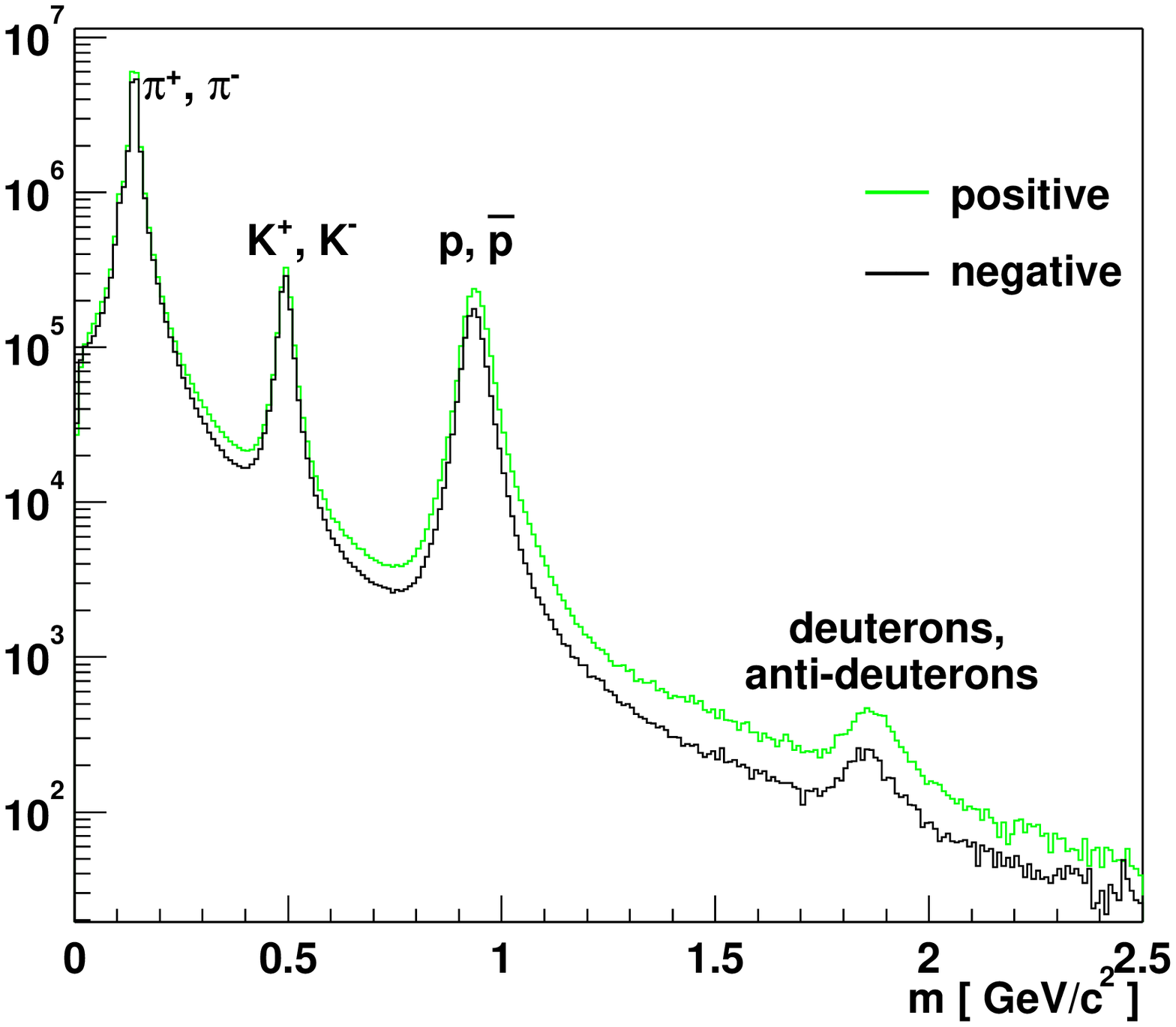}}
\caption[*]{Left: Scatter plot of $q \cdot (1-\beta)$ vs. $1/p$ for tracks in the central 
tracking arm. Right: Yield as a function of mass for positive and negative particles.}
\vspace{-1.0cm}
\end{center}
\end{figure}

\begin{figure}[!t]
\begin{center}
\centerline{\epsfxsize=2.0in\epsfbox{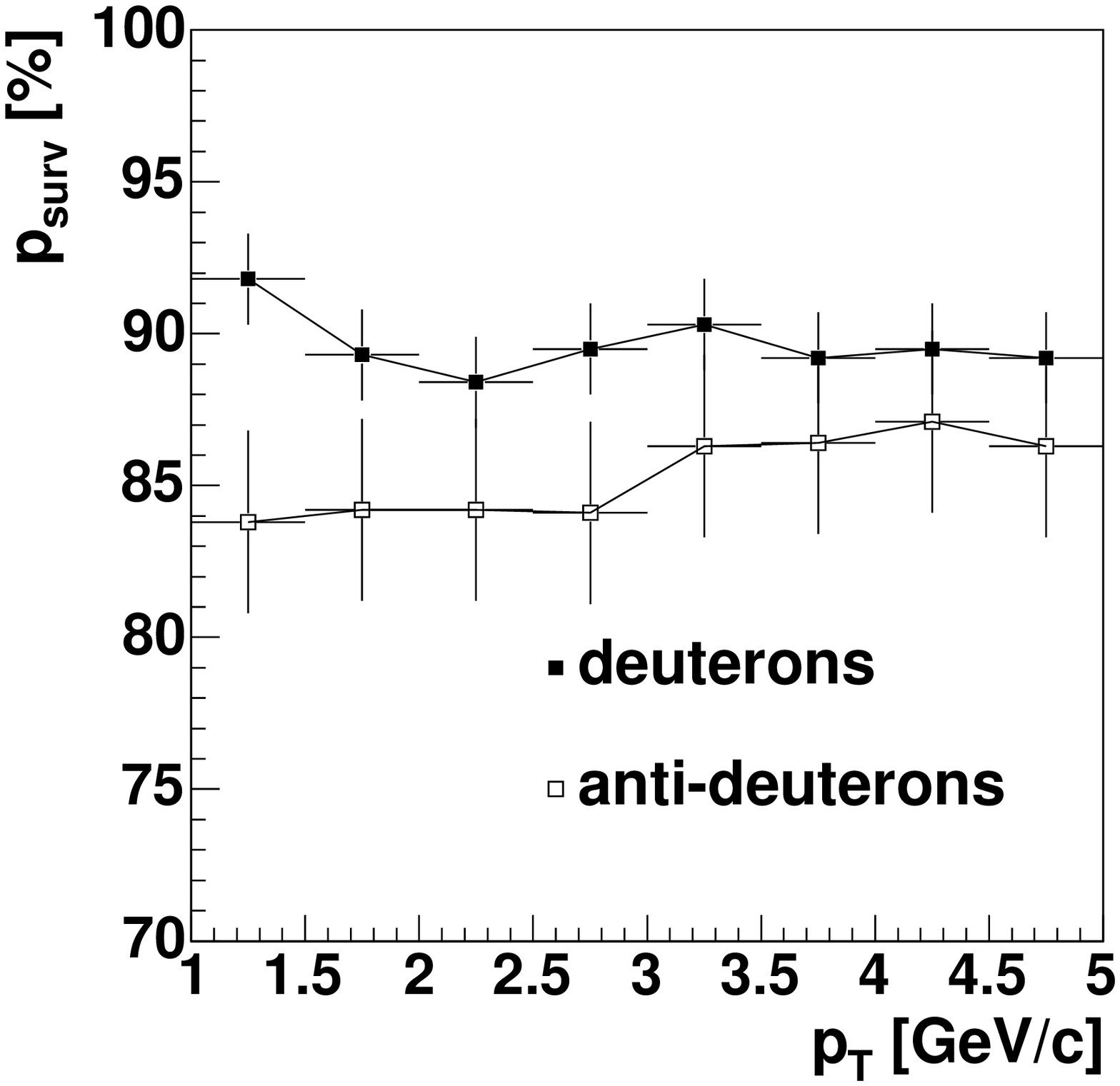} \epsfxsize=2.0in\epsfbox{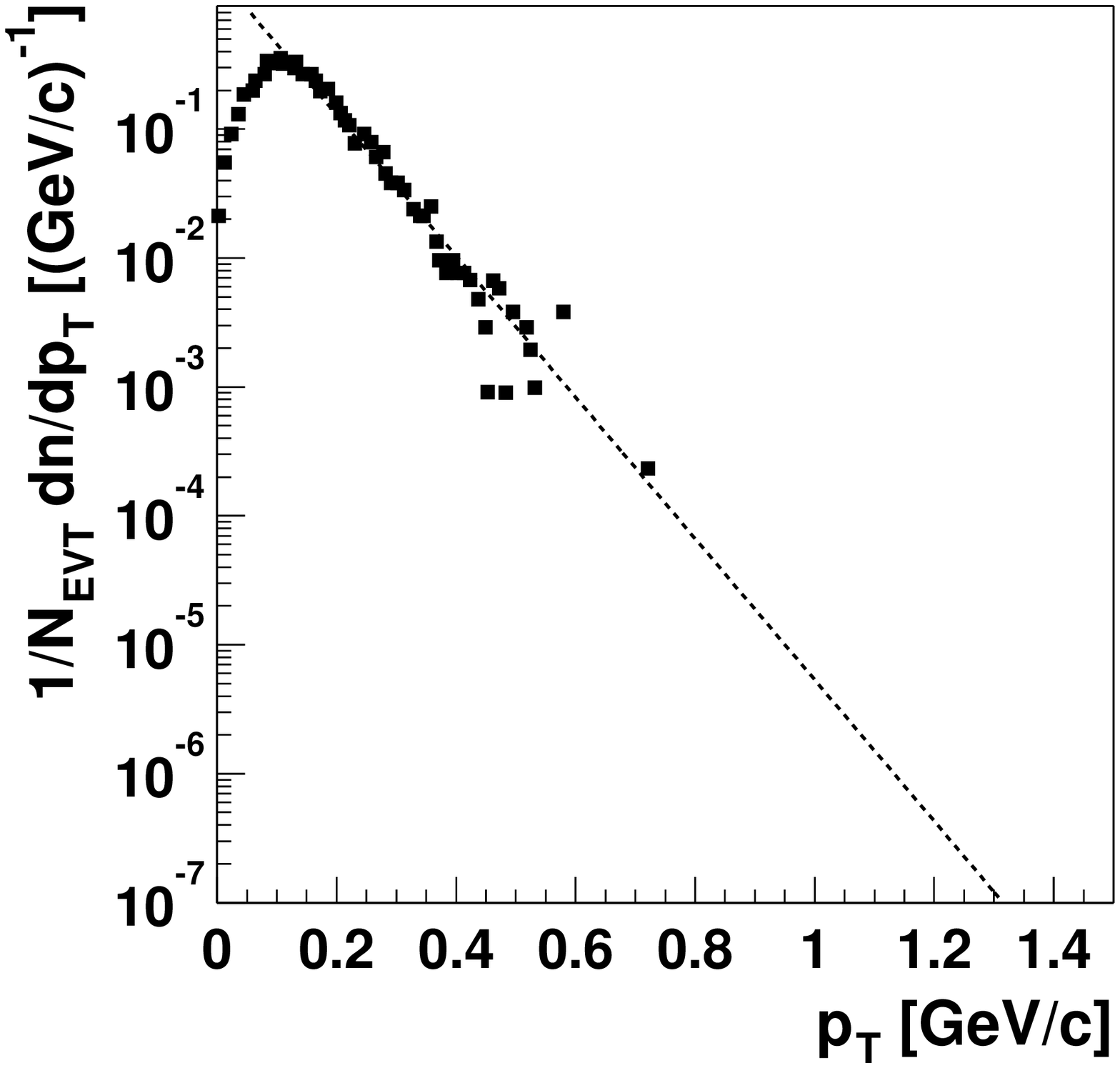}}
\caption[*]{Left: The probability as a function of transverse momentum that an (anti-)deuteron will reach the 
ToF without undergoing any inelastic hadronic interaction. The limited statistics in the GEANT simulation is 
partly responsible for the scattering of the points. 
Right: Transverse momentum distribution of knock-out deuterons from the detector material. The dotted line 
is an exponential fit\cite{RdR}.}
\vspace{-1.0cm}
\end{center}
\end{figure}

The information on what tracking media and corresponding materials lie between the primary vertex 
and the ToF was extracted from the PHENIX simulation program using 'geantinos', a virtual, 
non-interacting particle defined in GEANT. A total of 1000 geantinos were generated, originating 
from the primary vertex and hitting the ToF uniformly distributed over the surface. 

From the path lengths, $l_i$, and densities, $n_i$, of the different materials, 
and the corresponding cross sections $\sigma_i$, 
the probability that a particle should survive without suffering any 
hadronic interaction is given by
\begin{equation}
    \exp( - \sum_i \sigma_i n_i l_i ) \; .
\end{equation}

The experimental data on total cross sections for deuteron induced interactions are limited, and 
there are of course no data on anti-deuteron induced interactions. A direct parameterization 
of $\sigma(d+A)$ or $\sigma(\overline{d}+A)$ is therefore not possible. Instead, a 
method has been developed which calculates the (anti-)deuteron inelastic cross sections 
from the (anti-)proton and (anti-)neutron cross sections. For $\sigma_{pA}$, 
$\sigma_{\overline{p}A}$, $\sigma_{nA}$, and $\sigma_{\overline{n}A}$, the parameterizations 
in \cite{Moiseev} are used.  

The effective nucleon-nucleus cross section is calculated as the average of the neutron- and 
proton-nucleus cross sections, $\sigma_{NA} = (\sigma_{pA} + \sigma_{nA})/2$,
and similarly for anti-nucleons. Based on the geometry of a nuclear collision, the 
deuteron cross section is then written 
\begin{equation}
\sigma_{dA} = \left[  \sqrt{ \sigma_{NA} } + \Delta_{d}(A)  \right]^2 \; . 
\end{equation}
$\Delta_{d}(A)$ corresponds to the average difference in radius between a nucleon and a deuteron
and can thus be expected to be largely independent of nuclear mass number and collision energy. 
Comparisons with deuteron data\cite{Jaros:1977it} shows that a constant 
$\Delta_{d}(A) = 3.51 \pm 0.25$~mb$^{1/2}$ can be used. 



Using the parameterizations of $\sigma_{pA}$ and $\sigma_{\overline{p}A}$, the hadronic interaction
probabilities of (anti-)protons were first calculated from Eq.~1. This calculation was then compared with 
the interaction probabilities calculated from GEANT/FLUKA. The difference between the two methods was 
small. The final nucleon interaction probabilities were calculated as the mean of the two methods, and the 
difference was used as an estimate of the systematic error. 
The ratio of the hadronic interaction probabilities of (anti-)deuterons to those of 
(anti-)protons were then calculated using the cross sections from Eq.~2 in Eq.~1. 
These ratios depend on the size of the target; for (anti-)protons with momenta of 0.75 GeV/c they are 
$d/p=$~1.79, $\overline{d}/\overline{p}=$~1.34 for hydrogen targets and $d/p=$~1.43, 
$\overline{d}/\overline{p}=$~1.16 for copper targets, for example. The resulting ratios 
($d/p$, $\overline{d}/\overline{p}$) were applied to the nucleon interaction probabilities. 
The final hadronic survivabilities for deuterons and anti-deuterons are plotted in Figure~2 (left). 

The simulation of the experimental acceptance is done for single particles in the 
detector. The effect of high occupancy has been investigated by embedding simulated particles in 
real events. 
The result is corrections of 6.9$\pm$2.7\%, 19.8$\pm$7.1\%, and 2.2$\pm$3.1\% for 
min. bias (0-92\%), central (0-20\%), and non-central (20-92\%) collisions, respectively. The embedding 
corrections are largely independent of $p_T$.   

Finally, possible background sources must be investigated. For anti-deuterons, this is not a problem; any 
anti-deuterons seen in the detector must have been produced in the primary collision between 
the two gold nuclei. Deuterons, however, can be produced in secondary (knock-out) reactions between the produced 
particles and the detector material. This has been investigated by running simulated Au+Au events from Hijing 
through the GEANT simulation program. The rate of production of secondary deuterons, primarily from interactions 
in the beam-pipe and the PHENIX Multiplicity Vertex Detector, is indeed quite high, but, as can be seen in 
Figure~2 (right), 
the yield is concentrated at very low transverse momenta. It is estimated that the contribution from knock-out 
deuterons in the lowest $p_T$-bin (1.1$\leq p_T \leq$1.5~GeV/c) is less than 0.1\%. No 
corrections for background deuterons have therefore been applied. 

\begin{figure}[!b]
\begin{center}
\centerline{\epsfxsize=2.4in\epsfbox{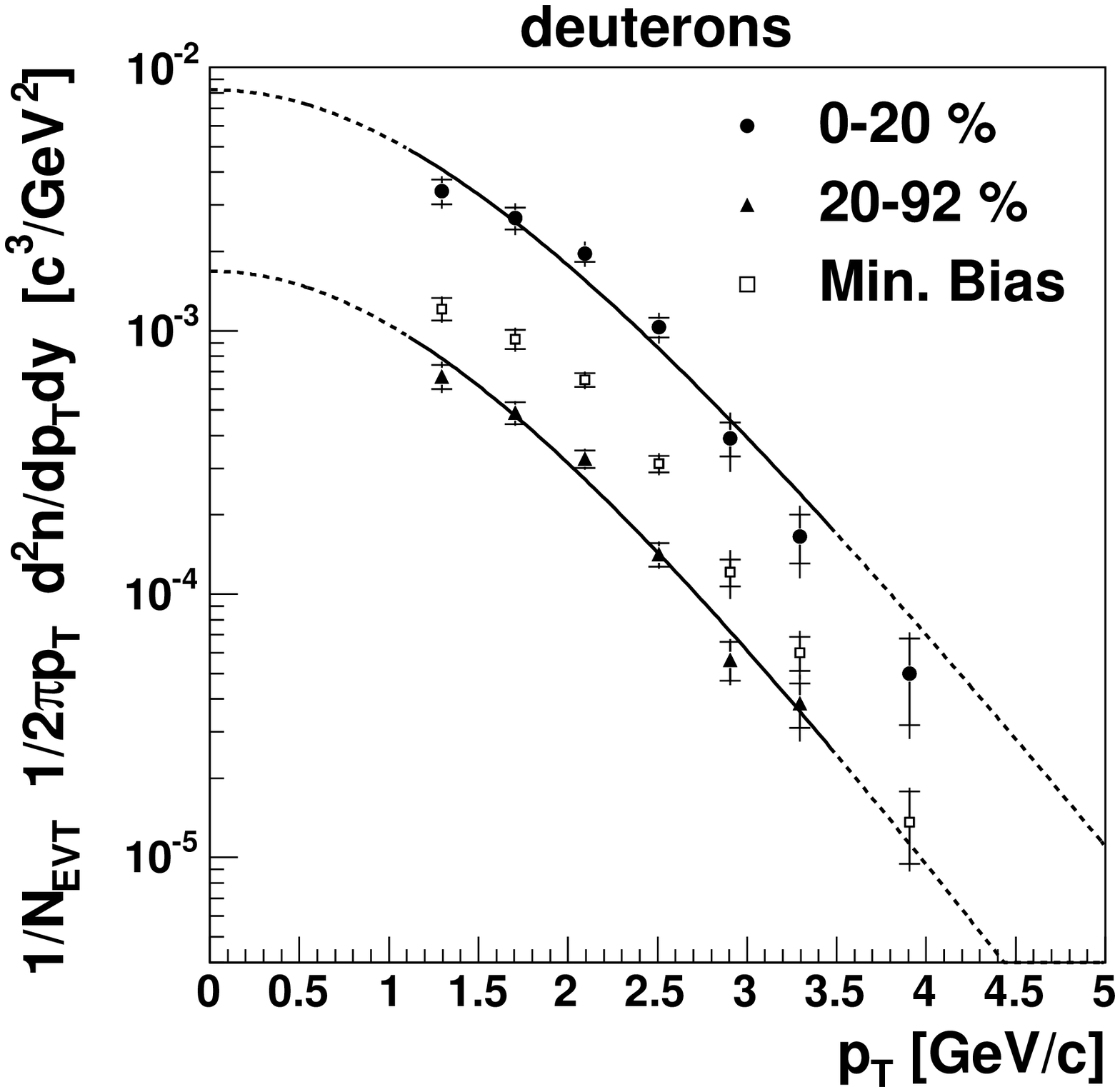} \epsfxsize=2.4in\epsfbox{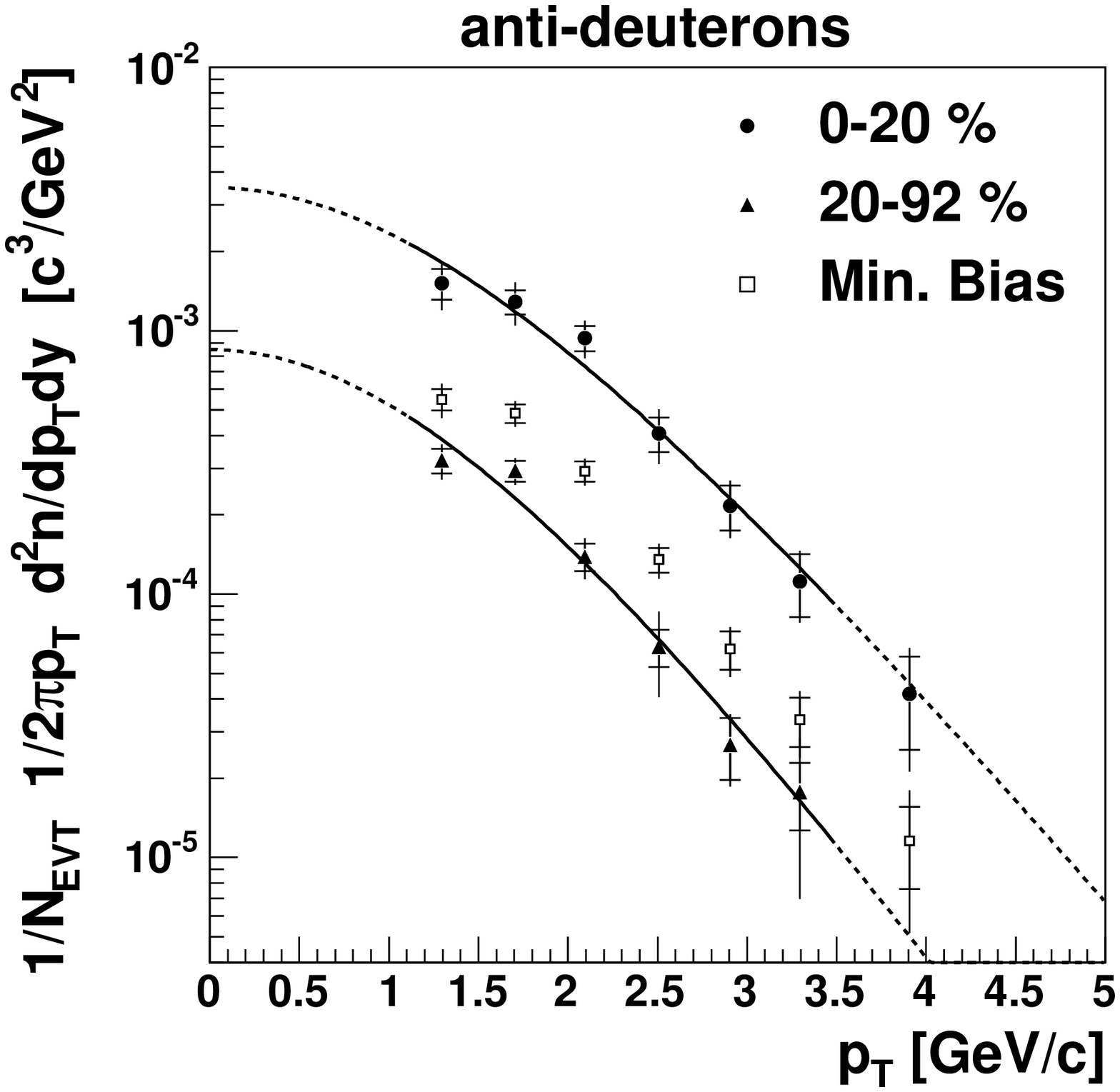}}
\caption[*]{Invariant yield of deuterons and anti-deuterons as a function of transverse momentum for 
three centrality selections. The error bars show the total error (quadratic sum of stat. and syst.) 
The horizontal bars show the statistical errors.}
\vspace{-1.0cm}
\end{center}
\end{figure}

The results after corrections are shown as normalized invariant yields in Figure~3. The spectra from the 
central and non-central samples have been fit to a Boltzmann distribution for a narrow interval in rapidity, 
\begin{equation}
\frac{d^2n}{p_T dp_T dy} = C \cdot m_T \cdot e^{-m_T/T} \; ,
\end{equation}
over the range 1.1~$\leq p_T \leq$~3.5~GeV/c. These fits give temperatures of $452\pm23$~MeV (0-20\%) and 
$421\pm21$~MeV (20-92\%) for deuterons and $475\pm37$~MeV (0-20\%) and 
$412\pm31$~MeV (20-92\%) for anti-deuterons. 

These temperatures are considerably higher than what has been observed for anti-deuterons produced in pp collisions 
at $\sqrt{s} =$~53~GeV. A fit to the combined data of \cite{Alper:1973my} in the range 
0.15~$\leq p_T \leq$~1.0 GeV/c gives $T = 110\pm26$~MeV. 

By extrapolating the Boltzmann fits (dashed curves in Fig.~3) the mean transverse momentum at mid-rapidity 
can be calculated. 
The $<p_T>$ of pions, kaons, and protons\cite{Adler:2003cb} are compared with that of 
\mbox{(anti-)}deuterons in 
Figure~4 (left). As can be seen, the $<p_T>$ of (anti-)deuterons is much higher than for the lighter particles. 
The results are well described by a linear function in the particle mass: 
\begin{equation}
<p_T> = <p_T>_0 + \beta \cdot m c \; .
\end{equation}
A fit gives $<p_T>_0 =$~358~MeV/c and $\beta =$~0.65. The large values of $<p_T>$ for \mbox{(anti-)}deuterons 
clearly indicate that they get a large fraction of their energy from the collective radial flow produced in 
central nucleus-nucleus collisions. This is what one would na\"{\i}vely expect from the trend seen for the 
lighter 
particles in Fig.~4. It is, however, difficult to see how the (anti-)deuterons with a binding energy of 
only 2.2~MeV can survive the intense hadronic rescattering normally assumed to produce the strong 
radial flow. 

\begin{figure}[!b]
\begin{center}
\centerline{\epsfxsize=2.4in\epsfbox{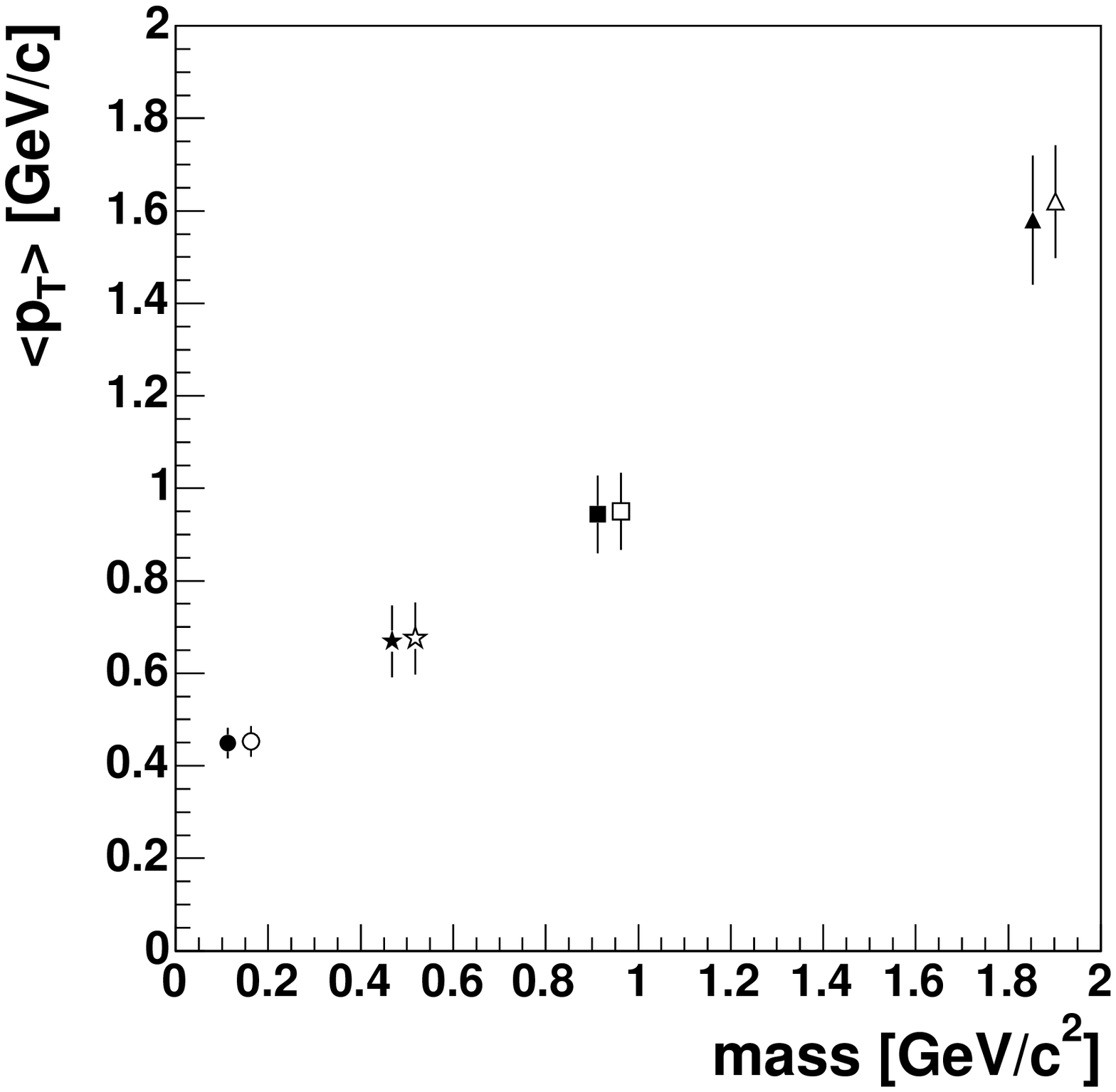} \epsfxsize=2.4in\epsfbox{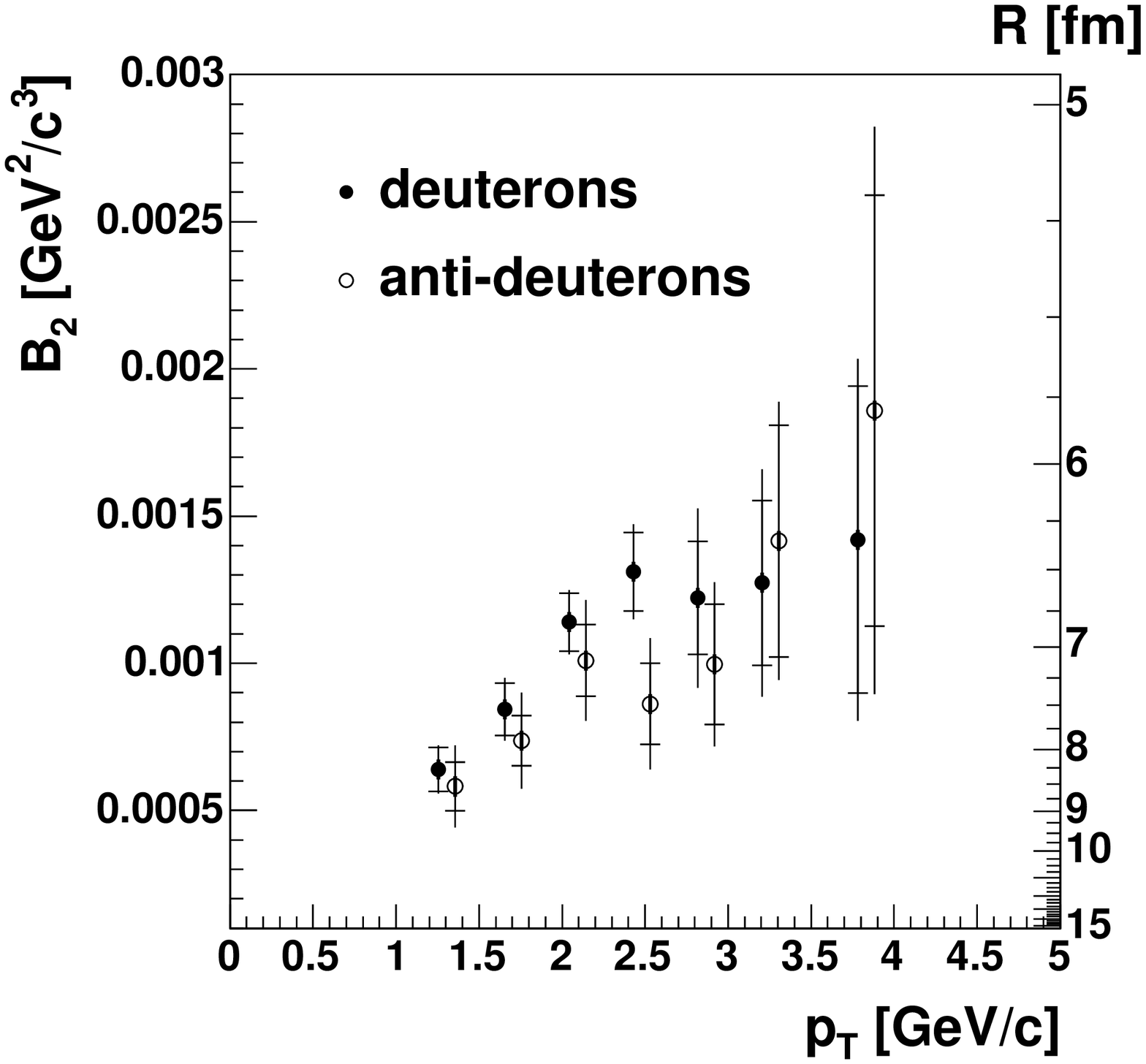}}
\caption[*]{Left: Mean transverse momentum vs. particle mass for pions, kaons, protons, and deuterons in 
central collisions. Solid and open symbols 
correspond to positively and negatively charged particles, respectively. To separate the two, the 
particle masses 
have been slightly shifted. The errors are in all cases dominated by the systematic errors. 
Right: The coalescence parameter $B_2$ as a function of $p_T$ for central collisions. The R values on the 
right-hand y-axis are from Eq.~\ref{eq:R}. The error bars are defined as in Fig.~3.}
\vspace{-1.0cm}
\end{center}
\end{figure}

The (anti-)deuterons produced at mid-rapidity in nuclear collisions at the current energy may be formed 
through neutron-proton coalescence at the late stages of the collisions. If one assumes that neutrons 
and protons (anti-neutrons and anti-protons) have the same phase space distribution, the invariant 
yield of deuterons (anti-deuterons) is given by 
\begin{equation}
E_d \frac{d^3n}{dp^3_d} = B_2 \left( E_p \frac{d^3n}{dp^3_p} \right)^2 
\label{eq:coalescence}
\end{equation}
where $B_2$ is the coalescence parameter for mass number A=2 and $p_p = p_d/2$. It should be noted that 
for $B_2 =$~constant, 
the coalescence model leads to approximately the same mean $p_T$ for protons and deuterons, 
$<p_T>_d \approx <p_T>_p$ (this relation is strictly true if 
the protons  have an exponential distribution in $p_T$, $dn/p_T dp_T \propto \exp(-p_T/T)$). 
If one assumes that the 
nucleons before freeze-out are uniformly distributed within a (static) sphere with radius R, $B_2$ can be 
related to R through (assuming $m_T \approx 2 m$)
\begin{equation}
R = \left( \frac{9 \pi^2}{2} \, \frac{(\hbar c)^3}{m_p \cdot B_2 \cdot c^5} \right)^{1/3} \; . 
\label{eq:R}
\end{equation}
$B_2$ for central collisions is plotted in Figure~4 (right) as a function of transverse momentum; 
the corresponding values of R are plotted on the right-hand y-axis. The (anti-)proton 
spectra from\cite{Adler:2003cb} are corrected for the contribution from weak decays before $B_2$ is 
calculated\cite{Adler:2004uy}. Eq.~6 gives a source radius near that of a gold nucleus. 
The increase of $B_2$ with $p_T$ and the observed large $<p_T>$ of (anti-)deuterons are, however, not 
consistent with a static source, and clearly point to the importance of collective flow. This estimate of 
$R$ can therefore only provide the correct order of magnitude. 

To summarize, PHENIX has measured the production of deuterons and anti-deuterons at mid-rapidity in gold-gold 
collisions at RHIC. 
The results show that the transverse momentum of the (anti-)deuterons primarily originates in the strong 
radial flow, despite the small deuteron binding energy. The invariant yields of \mbox{(anti-)}deuterons 
are not consistent with emission from a static source.

\end{document}